\documentclass[10pt,conference]{IEEEtran}
\IEEEoverridecommandlockouts
\usepackage{hyperref}
\usepackage{multirow}
\usepackage{booktabs}
\usepackage{bbm}
\usepackage{tabularx}

\usepackage{makecell}
\usepackage{pstricks}
\usepackage{bbding}
\usepackage{pifont}

\usepackage{amsmath} 
\usepackage{bm}      

\usepackage{cite}
\usepackage{amsmath,amssymb,amsfonts}
\usepackage{algorithmic}
\usepackage{graphicx}
\usepackage{CJKutf8}
\usepackage{textcomp}
\usepackage{xcolor}
\def\BibTeX{{\rm B\kern-.05em{\sc i\kern-.025em b}\kern-.08em
    T\kern-.1667em\lower.7ex\hbox{E}\kern-.125emX}}
\begin{document}
\begin{CJK}{UTF8}{gbsn}

\title{R-NeRF: Neural Radiance Fields for Modeling RIS-enabled Wireless Environments\\
 }


\author{\IEEEauthorblockN{Huiying Yang, Zihan Jin, Chenhao Wu, Rujing Xiong, Robert Caiming Qiu, Zenan Ling\textsuperscript{\dag}}
\IEEEauthorblockA{School of EIC, Huazhong University of Science and Technology, Wuhan 430074, China.\\ 
Ezhou Industrial Technology Research Institute, Huazhong University of Science and Technology, Wuhan 430074, China. \\ 
\textsuperscript{\dag}  Corresponding Author.
Email: lingzenan@hust.edu.cn}
\thanks{This work  was supported by the National Natural Science Foundation of China (via fund NSFC-12141107), the Guangxi Science and Technology Project (AB21196034), the Natural Science Foundation of Hubei Province (via fund 2024AFB074), the Key Research and Development Program of Hubei (2021BAA037), and the Guangdong Key Lab of Mathematical Foundations for Artificial Intelligence Open Fund (OFA00003). 
The datasets and code are available at: \href{https://github.com/HUSTGSNeRF/R-NeRF}{https://github.com/HUSTGSNeRF/R-NeRF}}
} 
\maketitle

\begin{abstract}

Recently, ray tracing has gained renewed interest with the advent of Reflective Intelligent Surfaces (RIS) technology, a key enabler of 6G wireless communications due to its capability of intelligent manipulation of electromagnetic waves.
However, accurately modeling RIS-enabled wireless environments poses significant challenges due to the complex variations caused by various environmental factors and the mobility of RISs.
In this paper, we propose a novel modeling approach using Neural Radiance Fields (NeRF) to characterize the dynamics of electromagnetic fields in such environments.
Our method utilizes NeRF-based ray tracing to intuitively capture and visualize the complex dynamics of signal propagation, effectively modeling the complete signal pathways from the transmitter to the RIS, and from the RIS to the receiver. This two-stage process accurately characterizes multiple complex transmission paths, enhancing our understanding of signal behavior in real-world scenarios. 
Our approach predicts the signal field for any specified RIS placement and receiver location, facilitating efficient RIS deployment. Experimental evaluations using both simulated and real-world data validate the significant benefits of our methodology.

\end{abstract}


\section{Introduction}

Reconfigurable Intelligent Surfaces (RISs) have recently emerged as a promising technology in wireless communications, especially in 6G development \cite{wu2019towards}. These surfaces are composed of programmable metasurface elements capable of dynamically manipulating electromagnetic waves, thereby customizing their electromagnetic properties to suit various incoming waves. This enables intelligent control over the wireless environment, leading to enhancements in signal strength, coverage, and overall network efficiency~\cite{di2020smart}.

Many existing works have concentrated on modeling
RIS-enabled wireless environments through electromagnetic properties~\cite{mi2023towards}, channel estimation~\cite{cheng2021joint}, and path loss models~\cite{tang2020wireless,huang2022reconfigurable}. 
Despite the advancements, these methods encounter practical challenges due to the assumption and simplification of channel models, and the complexities of optimization. 
In particular, such methods often necessitate repetitive adjustments when changing the receiver positions and typically overlook the effects of RIS positioning changes.
In case of practical radio propagation, ray tracing is a promising technique that has recently gained increased attention with the advent of RIS technology~\cite{degli2022reradiation}.
Studies such as~\cite{hao2024modeling,pyhtilae2023ray,huang2022novel}
have employed ray-tracing to model the channel environments influenced by the RIS. 
However, despite its detailed modeling of environmental interactions, conventional ray-tracing based methods are computationally intensive and often struggle to address the epistemic uncertainties in real-world scenarios.
While the Neural Radiance Fields (NeRF) technique has demonstrated remarkable performance in optical applications, recent researches have extended its application to electromagnetism\cite{zhao2023nerf2,yang2024codebook}.
Leveraging the advanced NeRF technique has opened new pathways in wireless communication modeling~\cite{orekondy2022winert,lu2024deep}. 
These developments enable accurately capturing the complex interactions of electromagnetic waves in dynamically changing environments.
Nevertheless, these studies only consider simple single-stage scenarios from signal transmission to reception, which differ from RIS-enabled environments. Such approaches may not be able to capture the role of the RIS in modifying the properties of the signal. Consequently, a more practical two-stage model is essential to adequately represent RIS-enabled wireless environments.


In this paper, we utilize NeRF-based ray tracing to capture the dynamic complexity of signal propagation and propose the R-NeRF model for RIS-enabled wireless environments.
Specifically, we introduce a subtly designed two-stage framework that properly tracks the entire transmission path. Each stage of our framework enables precise characterization of the electromagnetic signal propagation dynamics.
The R-NeRF method utilizes the coordinates of the RIS, the transmitter, and the receiver as inputs, enabling it to predict the signal at various receiver positions under different RIS placement strategies. 
This method adeptly captures the intricate effects of electromagnetic fields in real-world dynamic environments, including scattering, diffraction, and reflection. Our approach avoids assuming and simplifying channel models, allowing it to handle complex real environment with simplicity. 
Consequently, our method is capable of predicting the signal field in space for any given RIS placement, offering guidance for downstream task, such as efficiet RIS deployment\cite{hao2024modeling,huang2022novel}.


Accurately modeling these electromagnetic waves
poses a significant challenge, especially in complex real-world
scenarios where they undergo intricate changes due
to environmental influences. 
To overcome this challenge, we take inspiration from NeRF, originally designed for rendering complex optical volumetric scenes\cite{mildenhall2021nerf}. NeRF demonstrates remarkable capabilities in accurately representing and learning the behavior of light. Based on this concept, we integrate NeRF with ray tracing techniques, commonly utilized in electromagnetic physics, to model the propagation of electromagnetic signals.
This integration allows us to use the powerful representation capabilities of NeRF to precisely depict how each signal behaves under various environmental conditions, including radiation and transmission characteristics.
Experimental results validate the effectiveness of our approach
in accurately modeling electromagnetic signal propagation.
The main contributions of this paper can be summarized as follows.

\begin{itemize}
\item  We model a two-stage framework for simulating electromagnetic signal propagation in RIS-enabled wireless environments, allowing for accurate characterization of signal dynamics.
\item  By integrating NeRF-based ray tracing techniques with electromagnetic physics, we propose a novel R-NeRF method to model the signal field for any specified RIS placement and receiver location.
\item Experimental results from simulations and measured data demonstrate the effectiveness of our method.
\end{itemize}


\section{Methodology} \label{sec:PROBLEM ANALYSIS AND SYSTEM MODEL}
\begin{figure}
  \centering
  \includegraphics[width = 0.48\textwidth, trim=0 10 0 0, clip]{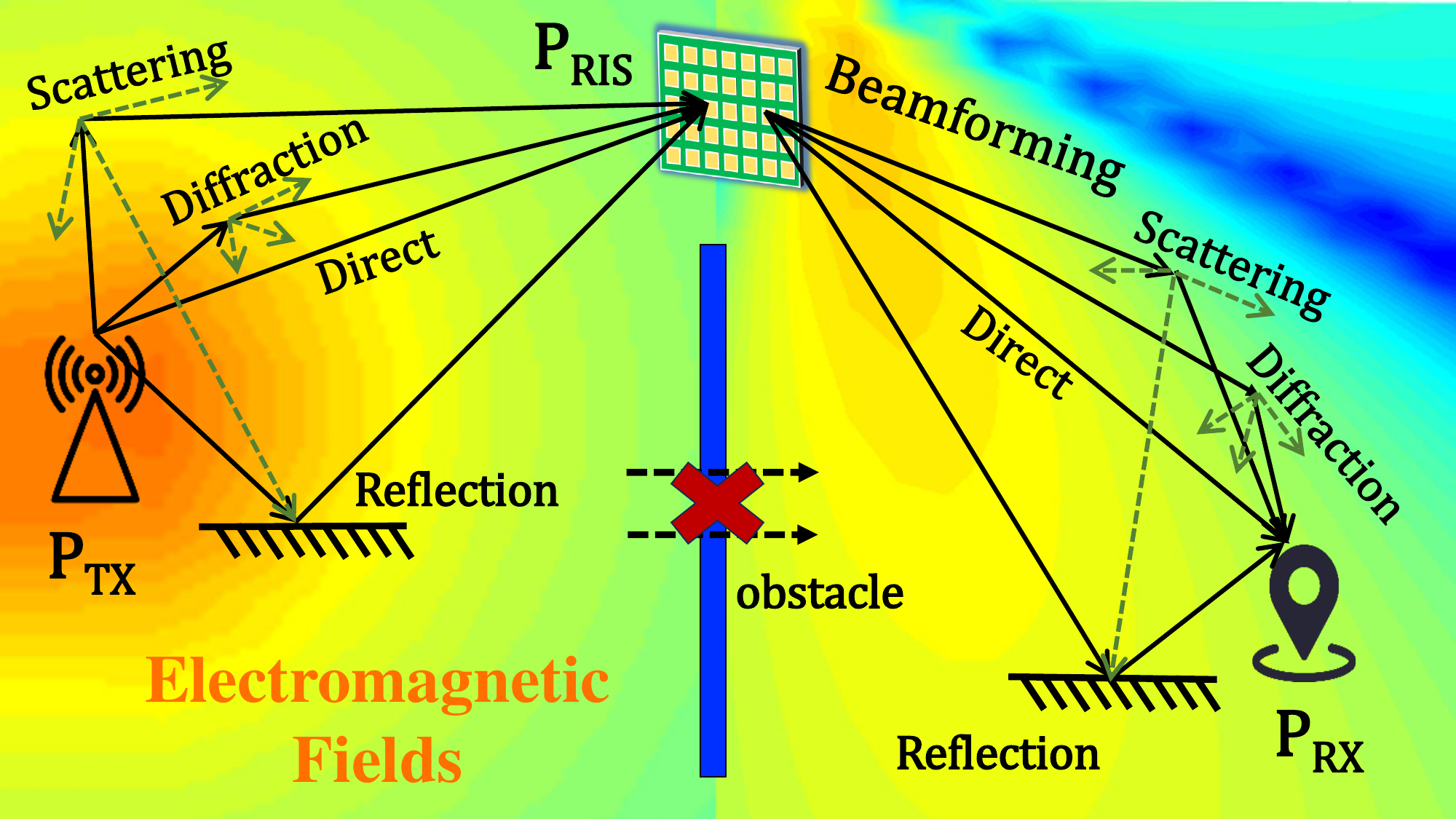}
  \caption{Illustration of the signal emission and transmission in RIS-enabled wireless environments，including its interactions with the environment through reflection, diffraction, scattering and etc.}
\label{fig:illustration enviroment.pdf}
\vspace{-0.6cm}
\end{figure}

We investigate a practical scenario involving a RIS-enabled communication system, as depicted in Fig. \ref{fig:illustration enviroment.pdf}. 
We consider the two-stage communication path from a single transmitter, through various placement of the RIS, to a receiver locates at different possible positions in the environment. Electromagnetic waves emitted from the transmitter undergo various changes due to the complex spatial interactions in the environment (scattering, reflection, diffraction, etc.). 
Our objective is to establish the implicit correlation between spatial coordinates and the received signal. By modeling the intricate dynamics of electromagnetic fields and capturing multiple transmission paths, we can predict the signal field for any given RIS placement and receiver location. This would enable valuable guidance for downstream tasks, such as efficient RIS deployment\cite{hao2024modeling,huang2022novel}.

\subsection{Preliminary: NeRF-based Ray Tracing} 

The problem is to model the intricate interactions between spatial position and the received signal. NeRF-based ray tracing technique employs the Fresnel principle, treating each emitted signal as comprised of multiple point sources to simulate signal propagation dynamics~\cite{zhao2023nerf2}. It includes three key components: ray tracing for modeling signal properties, rendering to accumulate signals from all point sources, and an implicit representation network to learn the relationships between these elements.


\textbf{Ray Tracing:}
The emitted signal consists of multiple point sources, each acting as an independent emitter, are denoted as voxels. 
We discretize the continuous signal spectrum by assuming the transmitter emits in $M$ distinct wave directions, and each ray is sampled at $N$ points along its trajectory, denoted as voxels $\mathbf{P_v}$. The ray received by the receiver in direction $\mathbf{d}$ can be modeled as
\begin{equation}
\mathbf{P_v}(t, \mathbf{d}) = \mathbf{o} + t\mathbf{d},
\label{eq:ray tracing}
\end{equation}
where $\mathbf{o}$ represents the receiver's coordinate, $\mathbf{d}$ is the view direction vector of the ray generated by, and $t$ indicates the distance information from the receiver.

For each voxel, we define $S_{mn}(\mathbf{P_v}, \mathbf{d})$ as the initial signal at the $n$-th voxel on the $m$-th ray, which is influenced by both the voxel's characteristics and the observation direction. As each voxel moves, the transmission factor $T_{mn}(\mathbf{P_v})$ captures environmental effects (RIS beamforming, reflection, scattering, etc.). This factor reflects the spatial impact on the voxel, independent of the observation direction. The mathematical relationships are expressed as follows:
\begin{equation}
S_{mn}(\mathbf{P_v}, \mathbf{d}) = A_{mn}(\mathbf{P_v}, \mathbf{d}) \cdot e^{i\alpha_{mn}(\mathbf{P_v}, \mathbf{d})},
\label{eq:S_mn}
\end{equation}
\begin{equation}
T_{mn}(\mathbf{P_v}) = \delta_{mn}(\mathbf{P_v}) \cdot e^{i\beta_{mn}(\mathbf{P_v})},
\label{eq:T_mn}
\end{equation}
where $A_{mn}(\mathbf{P_v}, \mathbf{d})$ and $\alpha_{mn}(\mathbf{P_v}, \mathbf{d})$ denote the initial amplitude and phase of the signal, respectively. The transmission amplitude $\delta_{mn}(\mathbf{P_v})$ and phase shift $\beta_{mn}(\mathbf{P_v})$ are influenced by the path-length difference as the signal travels from the voxel to the receiver, and are determined by:
\begin{equation}
\delta_{mn}(\mathbf{P_v}) = \frac{\rho}{t_{mn}(\mathbf{P_v})},\quad \beta_{mn}(\mathbf{P_v}) = \frac{2\pi\xi}{\lambda} \cdot t_{mn}(\mathbf{P_v}),
\end{equation}
where $\rho$ and $\xi$ are the amplitude and phase transmission coefficients, respectively, $\lambda$ is the wavelength of the transmitted signal, and $t_{mn}(\mathbf{P_v})$ indicates the distance traveled.
By discretizing the continuous signal spectrum into voxels and employing ray tracing techniques, we can precisely predict the signal characteristics of each voxel.

\textbf{Rendering:}
The total signal $R$ at the receiver is the sum of the signals contributed by all $N$ voxels on all $M$ rays:
\begin{equation}
R = \sum_{m=1}^M \sum_{n=1}^N S_{mn}(\mathbf{P_v}, \mathbf{d}) \cdot T_{mn}(\mathbf{P_v}).
\label{eq:sigle R}
\end{equation}

\begin{figure*}
  \centering
  \includegraphics[width=1.01\textwidth,trim=220 300 170 310, clip]{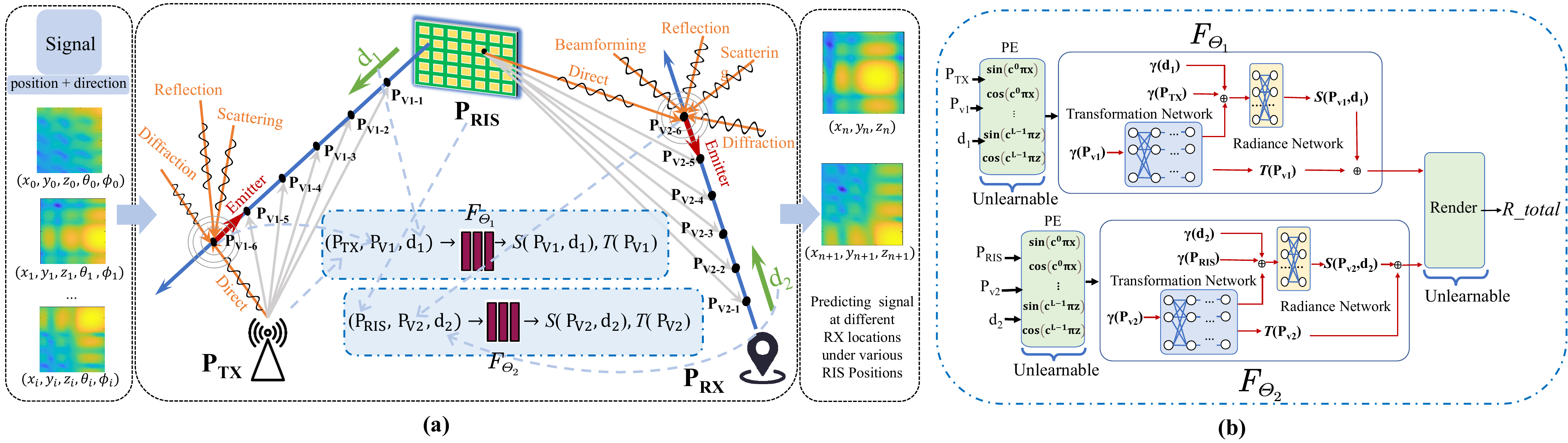}
  \caption{(a) The two-stage process of R-NeRF. Firstly, the ray emits from the TX to the RIS, represented by voxels $\mathbf{P_{v1-1}}$ to $\mathbf{P_{v1-6}}$. Each voxel along this path serves as a new transmitter, emitting the signal towards the RIS. Secondly, the signal continues its journey from the RIS to the RX, encountering voxels $\mathbf{P_{v2-1}}$ to $\mathbf{P_{v2-6}}$. At each voxel, the signal is changed by the intervening voxels, influencing its signal amplitude and phase upon reaching the destination. (b) Two-stage Neural network structure. We respectively use \( \mathbf{F_{\Theta_1}} \) and \( \mathbf{F_{\Theta_2}} \) to learn the signal characteristics of the TX-RIS stage and the RIS-RX stage, and employ rendering equations to obtain the total signal $R_{total}$.
 }
  \label{fig:similarity_comparison}
\end{figure*}

The total received signal at the receiver by summing the contributions from all voxels along multiple transmission paths. This modeling allows us to capture both the initial signal characteristics and the environmental influences on signal transmission.

\begin{figure*}
  \centering
  \includegraphics[width = 1.13\textwidth,trim=260 205 185 170, clip]{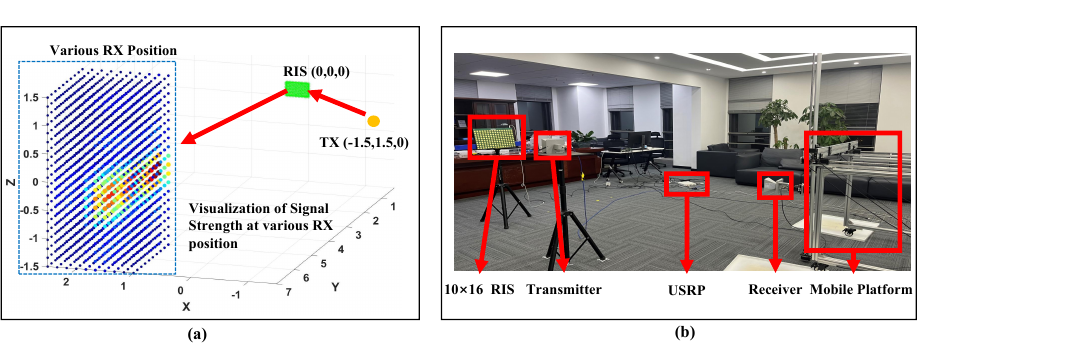}
  \caption{(a) The simulation scenario, (b) The experimental scenario.}
\label{fig:Simulated and measured scenarios.pdf}
\vspace{-0.3cm}
\end{figure*}

\textbf{Implicit Representation Network:}
NeRF-based methods use a neural network \( \mathbf{F_{\Theta}} \) to predict $S_{mn}(\mathbf{P_v}, \mathbf{d})$ and $T_{mn}(\mathbf{P_v})$, as described below:
\begin{equation}
\mathbf{F_{\Theta}} : (\mathbf{P_{TX}}, \mathbf{P_{v}}, \mathbf{d}) \rightarrow (S(\mathbf{P_{v}}, \mathbf{d}), T(\mathbf{P_{v}})),
\end{equation}
where $\mathbf{P_{TX}}$ is the position of the TX, and \( \Theta \) represents the network parameters.

\subsection{R-NeRF:Two-stage modeling Method}
We advances into a two-stage communication model for the RIS-enabled environment, as depicted in Fig. \ref{fig:similarity_comparison}(a). 
This two-stage approach ensures that the model captures the RIS's role in altering the signal's properties, which single-stage models might oversimplify or neglect. It reflects the RIS’s dual role in the communication pathway, both as a receiver that intercepts and processes the incoming signal and as a transmitter that actively shapes and directs the signal towards the receiver.
To be specific, the first stage models the transmission from the transmitter (TX) to RIS, treating the RIS as an intermediate receiver. The second stage considers the RIS as a new emitter, transmitting the processed signal towards the final receiver (RX).
Both stages employ the same principle of signal representation equation, facilitating a comprehensive analysis of signal propagation across diverse RIS placements and receiver locations. 
Assuming the RIS re-radiates in $M$ directions and samples $N$ points on each ray, and considering two direction vectors $\mathbf{d}_{\text{TX}\rightarrow\text{RIS}}$ and $\mathbf{d}_{\text{RIS}\rightarrow\text{RX}}$, the total received signal $R_{total}$ is the product of the two stages:

\begin{equation}
R_{total} = R_{\text{TX}\rightarrow\text{RIS}} \cdot R_{\text{RIS}\rightarrow\text{RX}},
\label{eq:R_total}
\end{equation}
\begin{equation}
R_{\text{TX}\rightarrow\text{RIS}} = \sum_{m=1}^M \sum_{n=1}^N S_{mn}(\mathbf{P_{v_1}}, \mathbf{d}_1) \cdot T_{mn}(\mathbf{P_{v_1}}),
\label{eq:TX-RIS}
\end{equation}
\begin{equation}
R_{\text{RIS}\rightarrow\text{RX}} = \sum_{m=1}^M \sum_{n=1}^N S_{mn}(\mathbf{P_{v_2}}, \mathbf{d}_2) \cdot T_{mn}(\mathbf{P_{v_2}}),
\label{eq:RIS-RX}
\end{equation}
where $\mathbf{P_{v_1}}$ and $\mathbf{P_{v_2}}$ are the positions of sampled points (voxels) along the rays in the transmission from the TX to the RIS and from the RIS to the RX, respectively. $\mathbf{d}_1$ and $\mathbf{d}_2$ are the directional vectors from the TX to the RIS and from the RIS to the RX, respectively.   
The two-stage modeling method provides a comprehensive framework for analyzing signal propagation in RIS-enabled communication systems. By treating the RIS as both an intermediate receiver and a new emitter in separate stages, we capture the intricate dynamics of signal transmission from the TX to the RIS and then from the RIS to the RX. This approach allows us to consider the impact of RIS-enabled paths on signal strength. 

\subsubsection{Two-stage Neural Network}
Based on our theoretical model of signal interactions in systems enabled by RIS, we now concentrate on implementing it through a predictive network.
The core objective of this network is to precisely estimate the signal characteristics at each voxel on each ray. The estimation process consists of estimating the signal's initial amplitude \( A_{mn}(\mathbf{P_{v}}, \mathbf{d}) \) and phase \( \alpha_{mn}(\mathbf{P_{v}}, \mathbf{d}) \), as well as their subsequent variations \( \delta_{mn}(\mathbf{P_{v}}) \) and \( \beta_{mn}(\mathbf{P_{v}}) \) during transmission.
In the TX-RIS stage, the network captures the characteristics of the signal as it moves from the TX to the RIS. A similar analysis is also conducted for the signal during the RIS to the RX.
The prediction model \( \mathbf{F_{\Theta}} \) that facilitates these assessments for both the transmission from the TX to the RIS and from the RIS to the RX are defined as: 
\begin{equation}
\mathbf{F_{\Theta_1}} : (\mathbf{P_{TX}}, \mathbf{P_{v_1}}, \mathbf{d}_1) \rightarrow (S(\mathbf{P_{v_1}}, \mathbf{d}_1), T(\mathbf{P_{v_1}})),
\end{equation}
\begin{equation}
\mathbf{F_{\Theta_2}} : (\mathbf{P_{RIS}}, \mathbf{P_{v_2}}, \mathbf{d}_2) \rightarrow (S(\mathbf{P_{v_2}}, \mathbf{d}_2), T(\mathbf{P_{v_2}})),
\end{equation}
where $S(\mathbf{P_{v_1}},\mathbf{d}_1)$, $T(\mathbf{P_{v_1}})$, $S(\mathbf{P_{v_2}},\mathbf{d}_2)$, $T(\mathbf{P_{v_2}})$ respectively represent the initial signal and transmission signal variations for the TX-RIS and RIS-RX paths.
The network's operation involves rendering the comprehensive signal \( R_{total} \), accumulating the predicted signal weights along the rays. 
By leveraging NeRF to model the intricate dynamics of electromagnetic fields and capture the multiple transmission paths, we achieve precise prediction of signal strength for various RIS placements and receiver locations.

\subsubsection{Neural Network Structure}
Our network comprises two main components: the unlearnable position encoding (PE) function and the learnable multi-layer perceptron (MLP) network, as depicted in Fig. \ref{fig:similarity_comparison}(b).

\textbf{Positional Encoding}: The PE function, $\gamma (\cdot)$, based on Fourier transform can map the normalised user coordinates to a high-dimensional space, which is shown below:
\begin{small}
\begin{equation}
\gamma(x)=(\sin(2^0\pi x),\cos(2^0\pi x),\cdots,\sin(2^{L-1}\pi x),\cos(2^{L-1}\pi x)).
\label{eq:position encoding}
\end{equation}\par 
\end{small}

\textbf{Multi-Layer Perceptron}: Our network architecture is specifically designed to address the unique challenges presented by RIS-enabled systems. It consists of two stages to model the signal: from the TX to the RIS and from the RIS to the RX. Each stage comprises two key components: the Transmission Network (TN) and the Radiation Network (RN), which learn the intricate signal variations and initial signal characteristics, respectively.
The TN is the first component in each stage, responsible for learning the transmission variations such as phase shifts and amplitude alterations as the signal traverses through the voxel space. In TX-RIS stage,
the TN network takes the voxel coordinate information as input and includes eight fully connected layers, each equipped with a ReLU activation function and 256 channels. It outputs the signal transmission parameters $T(\mathbf{P_{v_1}})$ as well as a 256-dimensional feature vector.
Subsequent to the TN, the RN receives the feature vector combined with the  $\mathbf{P_{TX}}$ and $\mathbf{P_{v_1}}$ to further process the initial signal characteristics $S(\mathbf{P_{v_1}}$, $\mathbf{d}_1)$. The RN consists of two fully connected layers with 256 and 128 channels, respectively, again employing the ReLU activation function. In the subsequent RIS-RX stage, we replicate this architectural paradigm. Consequently, we use the Eq. \eqref{eq:R_total}, \eqref{eq:TX-RIS}, \eqref{eq:RIS-RX} to render the final received signal.

\subsection{Model Training and Inference}
\subsubsection{Training}
After normalizing the data, we perform ray tracing separately from the RIS and the RX as origins. Emitting $M$ rays uniformly in their respective directions from the RIS and the RX, we identify $N$ voxel positions along each ray's trajectory. These voxels serve as points of interaction for the emitted signal. Positional Encoding (PE) is then applied to the ray directions, as well as to the coordinates of the voxels, TX, RIS, and RX, following Eq. \eqref{eq:position encoding}. This PE process effectively translates spatial information into a higher-dimensional space, capturing intricate spatial relationships and high-frequency functions. The encoded spatial information for each ray's direction and the corresponding voxel positions are subsequently fed into our neural network as input for training. Our neural network comprises both TN and RN, and after the rendering process, we obtain the predicted signal at the RX.
For the predicted signal strength $\hat{R}_i$ and the ground-truth signal strength $R_i$, we use MSE loss function to calculate the error: 
$L = \frac{1}{N} \sum_{i=1}^{N}(\hat{R}_i-R_i)^2$, where $N$ is the data number.

\begin{table}
\caption{Hyper-parameter settings}
    \centering
    \begin{tabular}{l|l|l}
         \hline
         Parameter &  Transmission Network & Radiation Network\\
         \hline 
         Dimension of PE & L=10 & L=10\\
         Dimension of input & 6L+3 &6L+3+6L+256\\
         Layer number & 8 & 2\\
         Batch size &128 & 128\\
         \hline
    \end{tabular}
    \label{tab:parameter settings}
    \vspace{-0.2cm}
\end{table}
\subsubsection{Inference}
Once the neural network model is trained, we utilize it for inference to predict the signal strength at the RX for unseen data. We start by normalizing the input data and using ray tracing to obtain voxel positions and ray directions, respectively. Then, we apply PE to encode spatial information into a higher-dimensional space. Next, the encoded information is fed into both the TN and RN. After rendering, the network outputs the predicted signal strength $\hat{R}_i$. Error calculation is performed by comparing the predicted strength with the ground-truth signal strength $R_i$ using a predefined loss function. 



\section{Experiment}
In this section, experiments on both simulated and measured data are constructed to evaluate the effectiveness of the proposed R-NeRF method. We use the Adam optimizer with an initial learning rate $10^{-3}$ to optimize neural network. Important parameter settings are outlined in Table~\ref{tab:parameter settings}. We compare our approach with several benchmarks, including $\text{NeRF}^2$\cite{zhao2023nerf2}, MRI\cite{Shin2014mri}, and MLP-based techniques. Specifically, the MRI method interpolates signal values at unsampled locations based on a rudimentary radio propagation model. We also conduct a series of ablation experiments to evaluate the impact of each component on the overall performance.

\begin{figure}
  \centering
  \includegraphics[width = 0.45\textwidth, trim=190 80 200 100, clip]{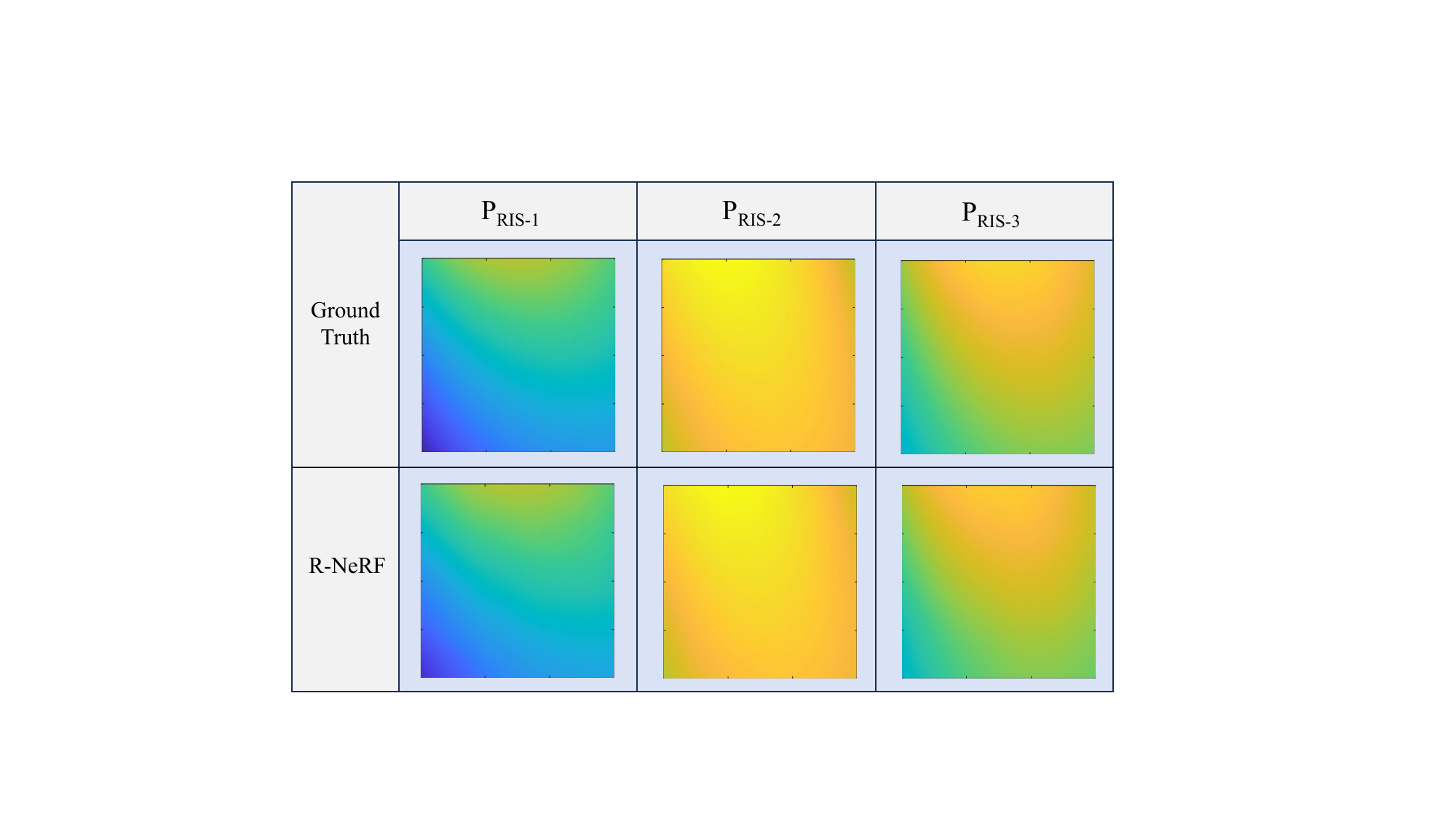}
  \caption{Visualization results of the ground-truth signal field (Top) v.s. R-NeRF predicted signal filed (Bottom) w.r.t different RIS positions: $\mathbf{P_{RIS-1}}$ (-0.063, 0, 0.8259),  $\mathbf{P_{RIS-2}}$ (0.4703, 0, -0.063), $\mathbf{P_{RIS-3}}$ (0.8259, 0, 0.8259).}
\label{fig:visual2.pdf}
\vspace{-0.6cm}
\end{figure}

\subsection{Dataset Construction}\label{subsec:Dataset Construction}
We construct datasets for both the simulated and the measured scenarios. The scenarios are shown in Fig.~\ref{fig:Simulated and measured scenarios.pdf}.\par
\subsubsection{Simulation Dataset}
We construct a simulated scenario, as depicted in Fig.~\ref{fig:Simulated and measured scenarios.pdf}(a), to captures the complex dynamics of an RIS-enabled environment within a designed three-dimensional space. This simulation, based on RIS modeling mechanisms~\cite{mi2023towards}, computes the signal strength at various RX positions under different RIS placements. We utilize these signal strength simulations as our ground-truth data. The simulation generated a dataset of 200,000 data points. The TX is positioned in the space at coordinates $(1.5, -1.5, 0)$. The RIS is variably positioned along the x and y axes, ranging from $-0.8$m to $0.8$m. The RX is strategically placed within a spatial range extending from $1$m to $2.5$m on the x-axis, $5$m to $7$m on the y-axis, and $-1.5$m to $1.5$m on the z-axis.

\subsubsection{Measured Dataset}
The experimental scenario is depicted in Fig.~\ref{fig:Simulated and measured scenarios.pdf}(b). The setup includes a RIS, a USRP device, and two horn antennas with the capability to transmit or receive signals in specific directions. The mobile platform offers an accessible space measuring $0.84$m$\times$ $0.6$m $\times$$0.6$m, establishing a grid of $15 × 7 × 5 = 525$ points. For the RIS, we uniformly select $16$ different positions in space. We employ a traversal algorithm to obtain the signal strength at a variety of RX locations for different RIS deployments~\cite{pei2021ris}. We utilize these signal strength measurements as our ground-truth data. \par

\subsection{Performance Evaluation}
\subsubsection{Evaluation Metrics}
Let \( E_{ri} \) represent the ground-truth data and \( E_{pi} \) denote the predicted data, where \(i = 1,\cdots, n \) and $n$ is
the data number. We employ three metrics to evaluate the predictive performance: (1) mean absolute error (MAE) \(\frac{1}{n} \sum_{i=1}^{n} |E_{ri} - E_{pi}|\), (2) median absolute error (MED) \(\text{median}\{|E_{ri} - E_{pi}|\}\), and (3) root mean squared error (RMSE) \(\sqrt{\frac{1}{n} \sum_{i=1}^{n} |E_{ri} - E_{pi}|^2}\).
\vspace{0.1cm}

\subsubsection{Visualization of signal field}
We visualize both the ground-truth signal field and the signal field predicted using the R-NeRF method. As shown in Fig. ~\ref{fig:visual2.pdf}, the results indicate that our method can forecast the signal strengths in the same space for different RIS placements, revealing variations in signal intensity. These findings offer guidance on the efficient RIS placement to enhance signal strength for the receivers, thereby improving communication efficiency. Furthermore, the close proximity between the predicted signal and the ground-truth values demonstrates the validity of our approach.

\subsubsection{Comparison with Benchmarks}
\begin{figure}
  \centering
  \includegraphics[width = 0.55\textwidth, trim=40 275 5 255, clip]{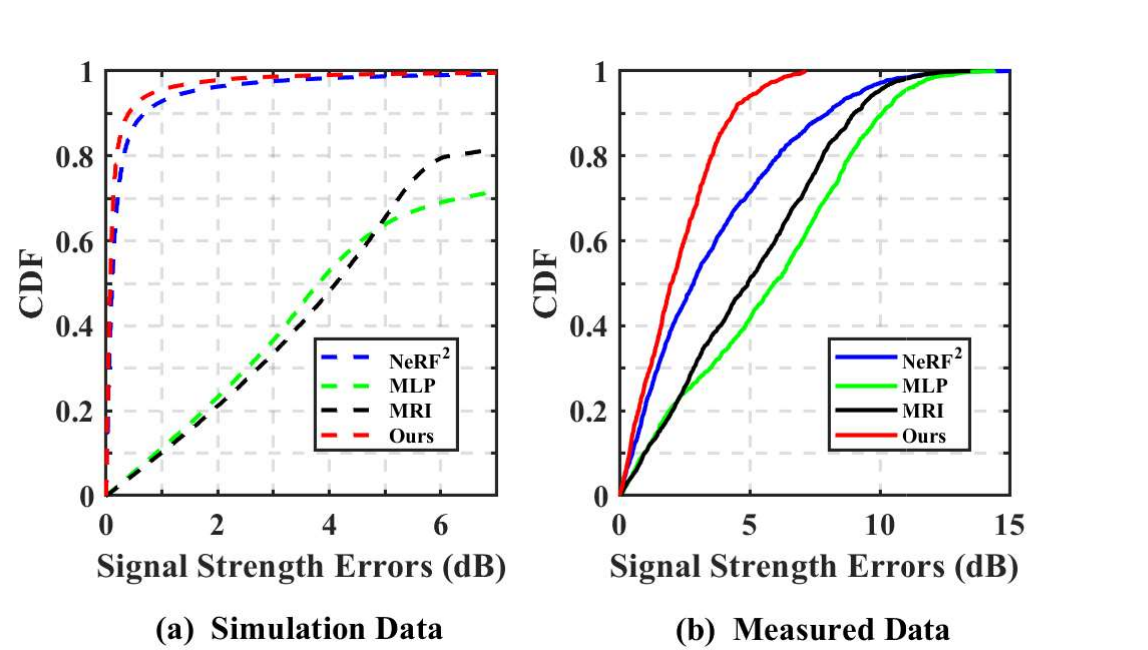}
  \caption{The CDF of the signal strength errors.}
\label{fig:CDF.pdf}
\vspace{-0.6cm}
\end{figure}

In comparative experiments, our proposed method surpasses the $\text{NeRF}^2$ method and significantly outperforms traditional MRI and MLP methods in predicting signal strength. The experimental results are shown in Fig.~\ref{fig:CDF.pdf}. The vertical axis of the graph represents the Cumulative Distribution Function (CDF), while the horizontal axis indicates signal errors. In the simulation data, the CDF curve of the our method rises more sharply and reaches closer to $1$ at a lower error value compared to other methods. 
In the measured data, the curve implies that it has the lowest error in signal strength prediction among all methods tested. Specifically, $94.13\%$ of the data in our method achieves a $5$dB error, the $\text{NeRF}^2$ method reaches $71.57\%$, the MRI method attains $51.34\%$, and the MLP method achieves $41.81\%$. These results underscore the effectiveness of our approach.

\par
\begin{figure}
  \centering
  \includegraphics[width = 0.5\textwidth, trim=30 280 30 250, clip]{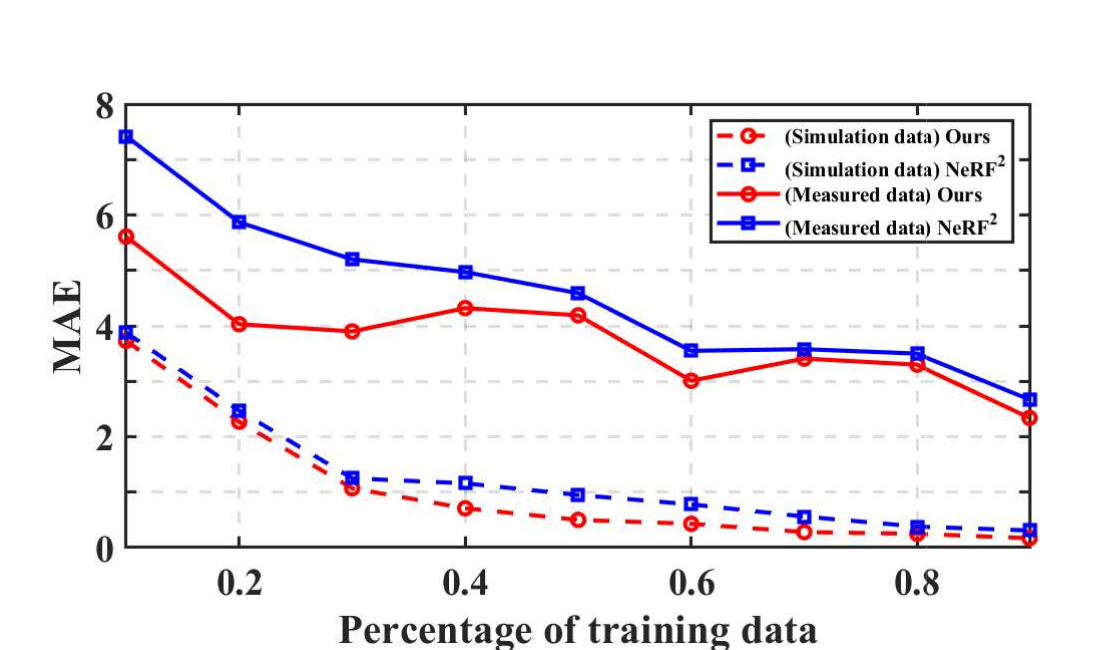}
  \caption{The MAE w.r.t the percentage of training data.}
\label{fig:MAE.odf}
\end{figure}

In addition to comparing the performance of different methods, we conducted experiments to explore how their effectiveness varies with the number of training and testing samples in the dataset. Fig.~\ref{fig:MAE.odf} presents comparative accuracy results between our method and $\text{NeRF}^2$ across varying proportions of training data. Our method consistently outperforms $\text{NeRF}^2$ on both simulated and measured datasets. Notably, with a small training dataset size, such as $10\%$ of the data for training and $90\%$ for testing, our method achieves a Mean Absolute Error (MAE) of $5.61$ dB, while $\text{NeRF}^2$ yields a MAE of $7.42$ dB. These findings suggest that our method delivers superior prediction performance even with limited datasets. Furthermore, as the dataset size increases, our method continues to outperform $\text{NeRF}^2$.
\subsubsection{Ablation Experiments}
To evaluate the individual contributions of specific components in our approach, we conduct ablation experiments. The results, showcased in Table~\ref{tab:ablation experiment}, illustrate that the integration of PE and ray tracing surpasses alternative methods. The experimental results show that when the ray tracing is not used, the MAE errors of both the simulated and measured data are above $5$ dB and the predictions are extremely poor, regardless of whether the PE is performed or not. Meanwhile, when using the ray tracing, the prediction can be further improved by using PE, and the MAE error is only $0.25$ dB on the simulated dataset. Notably, the inclusion of PE enriches the network's expressiveness by enhancing the dimension of the input coordinates. Concurrently, ray tracing enables networks to better capture feature information in complex propagation dynamics. As a result, ablation experiments demonstrate the effectiveness of the PE and ray tracing components in our method.\par



%
\section{Conclusion} 
In this paper, we have proposed a NeRF-based ray tracing method to model dynamic electromagnetic fields in RIS-enabled environments. This method can effectively capture complex propagation dynamics, accurately representing both signal emission and transmission across diverse positions. Our framework is structured into two key stages: transmission from the TX to the RIS, and from the RIS to the RX. This two-stage approach is pivotal for predicting signal field at different RX locations and strategically deploying RIS to improve communication efficiency. Empirical validation through simulated and measured data highlights the significant advantages of our proposed method. For the future work, we will investigate RIS-enabled scenarios with multiple transmitters and receivers.

\begin{table}
  \centering
  \caption{Ablation experiment}
  \label{tab:combined_performance}
  \begin{tabular}{cccccc}
    \toprule
     & Ray Tracing & PE & MAE & MED  & RMSE \\
    \midrule
    \multirow{4}{*}{Simulation data} & \multirow{2}{*}{NO}& \checkmark & 6.65 & 3.81  & 9.98 \\
    && $\times$ & 6.65 & 4.05  & 9.84 \\
     &\multirow{2}{*}{YES}& \checkmark & \textbf{0.25} & \textbf{0.07}  & \textbf{0.92} \\
    && $\times$ & 0.95 & 0.41  &2.29 \\
    \midrule
    \multirow{4}{*}{Measured data} & \multirow{2}{*}{NO}& \checkmark & 5.70 & 5.92  & 6.63 \\
    && $\times$ & 5.72 & 5.95  & 6.65 \\
     &\multirow{2}{*}{YES}& \checkmark & \textbf{3.3} & \textbf{2.56}  & \textbf{4.29}\\
     && $\times$ & 3.77 & 3.61 & 4.36 \\
    \bottomrule
  \end{tabular}
  \label{tab:ablation experiment}
  \vspace{-0.4cm}
\end{table}

\bibliographystyle{IEEEtran}
\bibliography{reference}

\begin{thebibliography}{10}
\providecommand{\url}[1]{#1}
\csname url@samestyle\endcsname
\providecommand{\newblock}{\relax}
\providecommand{\bibinfo}[2]{#2}
\providecommand{\BIBentrySTDinterwordspacing}{\spaceskip=0pt\relax}
\providecommand{\BIBentryALTinterwordstretchfactor}{4}
\providecommand{\BIBentryALTinterwordspacing}{\spaceskip=\fontdimen2\font plus
\BIBentryALTinterwordstretchfactor\fontdimen3\font minus \fontdimen4\font\relax}
\providecommand{\BIBforeignlanguage}[2]{{%
\expandafter\ifx\csname l@#1\endcsname\relax
\typeout{** WARNING: IEEEtran.bst: No hyphenation pattern has been}%
\typeout{** loaded for the language `#1'. Using the pattern for}%
\typeout{** the default language instead.}%
\else
\language=\csname l@#1\endcsname
\fi
#2}}
\providecommand{\BIBdecl}{\relax}
\BIBdecl

\bibitem{wu2019towards}
Q.~Wu and R.~Zhang, ``Towards smart and reconfigurable environment: Intelligent reflecting surface aided wireless network,'' \emph{IEEE Communications Magazine}, vol.~58, no.~1, pp. 106--112, 2019.

\bibitem{di2020smart}
M.~Di~Renzo, A.~Zappone \emph{et~al.}, ``Smart radio environments empowered by reconfigurable intelligent surfaces: How it works, state of research, and the road ahead,'' \emph{IEEE Journal on Selected Areas in Communications}, vol.~38, no.~11, pp. 2450--2525, 2020.

\bibitem{mi2023towards}
T.~Mi, J.~Zhang \emph{et~al.}, ``Towards analytical electromagnetic models for reconfigurable intelligent surfaces,'' \emph{IEEE Transactions on Wireless Communications}, 2023.

\bibitem{cheng2021joint}
X.~Cheng, Y.~Lin, and oters, ``Joint optimization for ris-assisted wireless communications: From physical and electromagnetic perspectives,'' \emph{IEEE Transactions on Communications}, vol.~70, no.~1, pp. 606--620, 2021.

\bibitem{tang2020wireless}
W.~Tang, M.~Z. Chen \emph{et~al.}, ``Wireless communications with reconfigurable intelligent surface: Path loss modeling and experimental measurement,'' \emph{IEEE Transactions on Wireless Communications}, vol.~20, no.~1, pp. 421--439, 2020.

\bibitem{huang2022reconfigurable}
J.~Huang, C.-X. Wang \emph{et~al.}, ``Reconfigurable intelligent surfaces: Channel characterization and modeling,'' \emph{Proceedings of the IEEE}, vol. 110, no.~9, pp. 1290--1311, 2022.

\bibitem{degli2022reradiation}
V.~Degli-Esposti, E.~M. Vitucci \emph{et~al.}, ``Reradiation and scattering from a reconfigurable intelligent surface: A general macroscopic model,'' \emph{IEEE Transactions on Antennas and Propagation}, vol.~70, no.~10, pp. 8691--8706, 2022.

\bibitem{hao2024modeling}
L.~Hao, S.~K. Vuyyuru \emph{et~al.}, ``Modeling ris from electromagnetic principles to communication systems--part ii: System-level simulation, ray tracing, and measurement,'' \emph{arXiv preprint arXiv:2403.13210}, 2024.

\bibitem{pyhtilae2023ray}
J.~Pyhtilae, J.~Kokkoniemi \emph{et~al.}, ``Ray tracing based radio channel modelling applied to ris,'' in \emph{WSA \& SCC 2023; 26th International ITG Workshop on Smart Antennas and 13th Conference on Systems, Communications, and Coding}.\hskip 1em plus 0.5em minus 0.4em\relax VDE, 2023, pp. 1--6.

\bibitem{huang2022novel}
J.~Huang, C.-X. Wang \emph{et~al.}, ``A novel ray tracing based 6g ris wireless channel model and ris deployment studies in indoor scenarios,'' in \emph{2022 IEEE 33rd Annual International Symposium on Personal, Indoor and Mobile Radio Communications (PIMRC)}.\hskip 1em plus 0.5em minus 0.4em\relax IEEE, 2022, pp. 884--889.

\bibitem{zhao2023nerf2}
X.~Zhao, Z.~An \emph{et~al.}, ``Nerf2: Neural radio-frequency radiance fields,'' in \emph{Proceedings of the 29th Annual International Conference on Mobile Computing and Networking}, 2023, pp. 1--15.

\bibitem{yang2024codebook}
H.~Yang, R.~Xiong \emph{et~al.}, ``Codebook configuration for ris-aided systems via implicit neural representations,'' in \emph{ICC 2024-IEEE International Conference on Communications}.\hskip 1em plus 0.5em minus 0.4em\relax IEEE, 2024, pp. 1449--1454.

\bibitem{orekondy2022winert}
T.~Orekondy, P.~Kumar \emph{et~al.}, ``Winert: Towards neural ray tracing for wireless channel modelling and differentiable simulations,'' in \emph{The Eleventh International Conference on Learning Representations}, 2022.

\bibitem{lu2024deep}
H.~Lu, C.~Vattheuer \emph{et~al.}, ``A deep learning framework for wireless radiation field reconstruction and channel prediction,'' \emph{arXiv preprint arXiv:2403.03241}, 2024.

\bibitem{mildenhall2021nerf}
B.~Mildenhall, P.~P. Srinivasan \emph{et~al.}, ``Nerf: Representing scenes as neural radiance fields for view synthesis,'' \emph{Communications of the ACM}, vol.~65, no.~1, pp. 99--106, 2021.

\bibitem{Shin2014mri}
H.~Shin, Y.~Chon \emph{et~al.}, ``Mri: Model-based radio interpolation for indoor war-walking,'' \emph{IEEE Transactions on Mobile Computing}, vol.~14, no.~6, pp. 1231--1244, 2014.

\bibitem{pei2021ris}
X.~Pei, H.~Yin \emph{et~al.}, ``Ris-aided wireless communications: Prototyping, adaptive beamforming, and indoor/outdoor field trials,'' \emph{IEEE Transactions on Communications}, vol.~69, no.~12, pp. 8627--8640, 2021.

\end{thebibliography}

\end{CJK}
\end{document}